\documentstyle[12pt]{article}

\input epsf.sty

\newcommand{\insertfig}[2]{\mbox{\epsfxsize=#1cm \epsfbox{#2.eps}}}

\newcommand{\Bx}{x_{\rm B}}
\newcommand{\ft}[2]{{\textstyle\frac{#1}{#2}}}

\newcommand{\Sym}{\mathop{\mbox{\large\bf S}}}

\begin{document}

\begin{titlepage}

\centerline{\large \bf Twist-three analysis of photon electroproduction
                       off pion.}

\vspace{15mm}

\centerline{\bf A.V. Belitsky$^a$, D. M\"uller$^{b}$,
                A. Kirchner$^b$, A. Sch\"afer$^b$}

\vspace{10mm}

\centerline{\it $^a$C.N.\ Yang Institute for Theoretical Physics}
\centerline{\it State University of New York at Stony Brook}
\centerline{\it NY 11794-3840, Stony Brook, USA}

\vspace{5mm}

\centerline{\it $^b$Institut f\"ur Theoretische Physik,
                Universit\"at Regensburg}
\centerline{\it D-93040 Regensburg, Germany}

\vspace{15mm}

\centerline{\bf Abstract}

\vspace{0.5cm}

\noindent

We study twist-three effects in spin, charge, and azimuthal asymmetries in
deeply virtual Compton scattering on a spin-zero target. Contributions which
are power suppressed in $1/Q$ generate a new azimuthal angle dependence of
the cross section which is not present in the leading twist results. On the
other hand the leading twist terms are not modified by the twist three
contributions. They may get corrected at twist four level, however. In the
Wandzura-Wilczek approximation these new terms in the Fourier expansion with
respect to the azimuthal angle are entirely determined by the twist-two
skewed parton distributions. We also discuss more general issues like the
general form of the angular dependence of the differential cross section,
validity of factorization at twist-three level, and a relation of skewed
parton distributions to spectral functions.

\vspace{5cm}


\vspace{0.5cm}
\end{titlepage}

\section{Introduction.}

Hard electroproduction of a photon, i.e.\ deeply virtual Compton scattering
(DVCS) \cite{MueRobGeyDitHor94,Ji97,Rad97}, measures non-forward matrix 
elements $\langle P_2| {\cal O}_i |P_1 \rangle$ of quark and/or gluon 
non-local composite operators ${\cal O}_i$. The former possesses a more 
diverse structure than processes with forward kinematics e.g.\ arising in 
conventional inclusive reactions. Fourier transforms of $\langle P_2| 
{\cal O}_i |P_1 \rangle$ define a new type of hadron characteristics --- 
generalized parton distributions (GPDs). Real photon electroproduction on 
nucleon targets has already been studied by DESY experiments: for moderate 
Bjorken varaible $\Bx$ by HERMES \cite{Ama00} and for small $\Bx$ by ZEUS 
and H1 collaborations \cite{Fav00}. Since these new functions are poorly 
known, our primary goal is to constrain diverse models from experimental 
data. To extract GPDs from future experiments \cite{Jlab} one has to address 
the problem of appropriate observables. The four-fold differential cross 
section for $e (k) h (P_1) \to e (k^\prime) h (P_2) \gamma (q_2)$
\begin{eqnarray}
\label{WQ}
\frac{d\sigma}{d\Bx dy d|\Delta^2| d\phi}
=
\frac{\alpha^3  \Bx y } { 8 \, \pi \,  {\cal Q}^2}
\left( 1 + \frac{4 M^2 \Bx^2}{{\cal Q}^2} \right)^{-1/2}
\left| \frac{\cal T}{e^3} \right|^2 
\end{eqnarray}
 depends on\ the Bjorken variable
$\Bx = - q_1^2/(2 P_1\cdot q_1)$ (with $q_1 = k - k'$ and $q_1^2 = - {\cal
Q}^2$), the momentum transfer $\Delta^2 = (P_2 - P_1)^2$, the fraction of the
lepton energy loss $y = P_1\cdot q_1/P_1\cdot k$ and the azimuthal angle
$\phi$ between lepton and hadron scattering planes. The amplitude ${\cal T}$
is a sum of the virtual Compton scattering (VCS) amplitude, ${\cal T}_{VCS}$, 
and the Bethe-Heitler (BH) amplitude, ${\cal T}_{BH}$, since they have the 
same initial and final states. Fortunately, the interference term, ${\cal I} 
\equiv {\cal T}_{VCS} {\cal T}_{BH}^\ast + {\cal T}_{VCS}^\ast {\cal T}_{BH}$, 
in $|{\cal T}|^2$ may be isolated by measuring various asymmetries like the 
lepton charge asymmetry, the hadron/lepton spin asymmetry and the azimuthal 
angle asymmetry \cite{Ji97,GouDiePirRal97,BelMueNieSch00}. Since ${\cal I}$ 
is linear in the DVCS amplitude, one can extract its real or imaginary part 
from an observable defined in an appropriate manner. Thus, one directly 
accesses a specific linear combination of GPDs convoluted with the real 
or imaginary part of a hard scattering amplitude.

The theoretical description of DVCS amplitudes is simplified for small
values of $\Bx$. This region is relevant for the HERA collider experiments.
For small $\Bx$ one naively assumes that GPDs are essentially determined by
the forward parton densities\footnote{The only free parameter is the
renormalization group scale where this identification is done.} multiplied
by partonic form factors. The momentum transfer dependence of GPDs may be
parametrized by a hadron form factor \cite{FraFreStr97} or for larger 
values it can be computed within QCD \cite{BalKuc00}. For larger values 
of $\Bx$ the `exclusive' domain of GPDs becomes important. Then their 
functional form cannot be approximated by conventional densities. Rather 
one has to rely on models or plausible parametrizations \cite{BelMueNieSch00}. 
Such predictions can be unstable under QCD radiative- and/or higher twists 
corrections. Therefore both issues deserve a detailed investigation. Although 
the evolution effects are quite moderate (of order $10-15\%$
\cite{BelMulNieSch99-2}), the next-to-leading order corrections in the
coefficient functions have a significant impact on the handbag approximation
and can be of order $50\%$ in the valence region \cite{BelMulNieSch99}. On
the other hand the exploration of power suppressed contributions has just
begun. Recent studies explored twist-three GPDs in the DVCS amplitudes
\cite{AniPirTer00}-\cite{RadWei00} and relations of twist-three GPDs to
already discussed leading twist functions \cite{BelMul00}-\cite{RadWei00}
and interaction dependent antiquark-gluon-quark correlations
\cite{BelMul00}. A natural next step is to study power suppressed
contributions in the coefficients of the Fourier expansion of the
differential cross section. For the sake of simplicity and clarity we
consider a pion target, $h = \pi$. While barely a target for an 
experiment, (pseudo) scalar particles are instructive for theoretical
reasons. They avoid difficulties accompanying particles with non-zero spin.
Our goal is thus to gain insights into gross features of twist-three effects 
in physical observables which are expected to hold true also for DVCS on 
higher spin targets. Presently, we discuss issues of the azimuthal angle 
dependence of the differential cross section (\ref{WQ}) with contributions 
from twist-three GPDs and qualitative estimates of higher twist effects.

The paper is organized as follows. In Section \ref{Sec-AziAngDep} we
calculate the azimuthal angle dependence of the differential cross
section in terms of the hadronic amplitudes, which appear in the DVCS
tensor for a spin-zero target. We derive constraints between the
angular moments of the amplitude squared. In Section \ref{Sec-Pre} we
evaluate the twist-three predictions based on perturbative leading
order calculations. In Section \ref{Sec-DD} we give a consistent
definition of spectral functions, so-called double distributions, and
derive their relationship to GPDs. After giving a few numerical
estimates in Section \ref{Sec-NumEst}, we summarize.

\section{Angular dependence of the cross section.}
\label{Sec-AziAngDep}

As we observed before, the total electroproduction amplitude of a real
photon consists of two terms, ${\cal T}_{VCS}$, we are interested in,
and the contaminating ${\cal T}_{BH}$. ${\cal T}_{BH}$ is purely real
and arises from a contraction of the leptonic tensor, $L_{\mu\nu} =
\bar u (k^\prime, \lambda^\prime) \left[ \gamma_\mu (\slash\!\!\! k -
  \slash\!\!\!\!\Delta)^{-1} \gamma_\nu + \gamma_\nu (\slash\!\!\!
  k^\prime + \slash\!\!\!\!\Delta)^{-1} \gamma_\mu \right] u(k,
\lambda)$, with a hadronic electromagnetic current, which for the pion
target reads $J_\mu = P_\mu F (\Delta^2)$. It is parametrized by the
electromagnetic form factor $F (\Delta^2)$ described by a simple
monopole form $\left( 1 - \Delta^2/m_V^2 \right)^{-1}$ with $m_V^2
\approx 0.46 \ \mbox{GeV}^2$.  Similarly, the DVCS amplitude is expressed by
${\cal T}_{VCS} = \pm \frac{e^3}{{\cal Q}^2} T_{\mu\nu}
\epsilon^\ast_\mu (q_2) \bar u (k^\prime) \gamma_\nu u(k)$, where the
sign depends on the beam charge, $\pm$ for $e^{\pm}$. The hadronic
tensor is given by
\begin{eqnarray}
\label{def-HadTen}
T_{\mu\nu} (q, P, \Delta) =
i \int dx {\rm e}^{i x \cdot q}
\langle P_2 | T j_\mu (x/2) j_\nu (-x/2) | P_1 \rangle.
\end{eqnarray}
It depends on the momenta $q = (q_1 + q_2)/2$, $P = P_1 + P_2$ and
$\Delta = P_2 - P_1$ which can be reexpressed in terms of the
Lorentz invariants $\xi = Q^2/P\cdot q$, $\eta = \Delta\cdot q /
P\cdot q$, $Q^2 = -q^2 = - \frac{1}{4}(q_1 + q_2)^2$ and $\Delta^2$.
For the DVCS kinematics, the first two scaling variables are related
to each other by $\eta = -\xi \left(1 - \frac{\Delta^2}{4 Q^2}\right)
\approx - \xi$. The `experimental' variables referred to in the
introduction are expressed by the present ones to twist-four accuracy
via, see e.g.\ \cite{BelMueNieSch00}, $Q^2 \approx \frac{1}{2} {\cal
  Q}^2$, $\xi \approx \Bx/(2 - \Bx)$. There are five independent
kinematical structures in Eq.\ (\ref{def-HadTen}) for a general two
photon process on a (pseudo) scalar target:
\begin{eqnarray}
\label{T-gen}
T_{\mu\nu}(q, P, \Delta) &=&
- {\cal P}_{\mu\sigma} g^{\sigma\tau} {\cal P}_{\tau\nu}
F_1
+ \frac{{\cal P}_{\mu\sigma} P^\sigma P^\tau {\cal P}_{\tau\nu}}{2 P \cdot q}
F_2
+ \frac{{\cal P}_{\mu\sigma} (P^\sigma \Delta^{\perp\tau}
+ \Delta^{\perp\sigma} P^\tau) {\cal P}_{\tau\nu}}{2 P \cdot q}
F_3
\nonumber\\
&&+ \frac{{\cal P}_{\mu\sigma}
(P^\sigma \Delta^{\perp\tau} - \Delta^{\perp\sigma} P^\tau)
{\cal P}_{\tau\nu}}{2 P \cdot q}
F_4
+ \frac{{\cal P}_{\mu\sigma}
\Delta^{\perp\sigma} \Delta^{\perp\tau} {\cal P}_{\tau\nu}}{M^2}
F_5    .
\end{eqnarray}
Here current conservation is ensured by means of the projector 
${\cal P}_{\mu\nu} = g_{\mu\nu} - q_{1 \mu} q_{2 \nu}/ q_1 \cdot q_2$.
The transverse component of the momentum transfer is $\Delta^\perp_\mu 
\equiv \Delta_\mu - \eta P_\mu$. We scaled the kinematical factors in 
such a way, that all five dimensionless scalar amplitudes change only 
logarithmically if quark masses are set to zero. Contracting the hadronic 
tensor with the leptonic current we can conclude that the amplitudes $F_1$, 
$F_2$ and $F_5$ are the leading contributions in ${\cal Q}$, while $F_3$ 
and $F_4$ are suppressed by the powers $1/{\cal Q}$ and $1/{\cal Q}^2$,
respectively. Note also that the Bose symmetry requires $F_1$, $F_2$,
$F_4$, $F_5$ to be even and $F_3$ to be odd in $\eta$. For the case at
hand the outgoing photon is real. Then the on-shell condition $q_2^2 =
0$ forces ${\cal P}_{\mu\sigma} \Delta_\sigma$ to be proportional to
$q_{2\,\mu}$.  When this term is contracted with the leptonic part its
contribution vanishes. Only three contributions are left then
\begin{eqnarray}
\label{T-DVCS}
\left. T_{\mu\nu}(q, P, \Delta) \right|_{q_2^2 = 0}
\!\!\! &=&\!\!\! - {\cal P}_{\mu\sigma} g_{\sigma\tau} {\cal P}_{\tau\nu}
\left(
T_1 
+ \frac{(1 - \xi^2)(\Delta^2 - \Delta^2_{\rm min})} {4\xi M^2} T_3 
\right)
\nonumber\\
&&\!\!\!+ 
\frac{{\cal P}_{\mu\sigma} P_\sigma P_\tau {\cal P}_{\tau\nu}}{2 P \cdot q}
\left(
T_2 + \frac{\Delta^2}{4 M^2} T_3 \right)
+ \frac{{\cal P}_{\mu\sigma} P_\sigma \Delta^\perp_\tau
{\cal P}_{\tau\nu}}{2 M^2} T_3 .
\end{eqnarray}
The new form factors are parametrized in such a way that gluonic
transversity only contributes to $T_3$ at ${\cal O}(\alpha_s)$. We also
dropped $T_3/{\cal Q}^2$ contributions in the first Lorentz structure.
The new amplitudes are parametrized by Compton form factors, which are
related to the old ones by the following set of equations
\begin{eqnarray*}
&& T_1 = 
\left( 
F_1 - \frac{(1 - \xi^2)(\Delta^2 - \Delta^2_{\rm min})}{2 M^2} F_5 
\right)_{|{\eta = - \xi}},
\qquad
T_2 =
\left(
F_2 + \xi F_3 - \xi F_4 - \xi \frac{\Delta^2}{2 M^2} F_5 
\right)_{|\eta = - \xi},
\nonumber\\
&& T_3 = \left. 2 \xi F_5 \right|_{\eta = - \xi} ,
\end{eqnarray*}
We again neglected twist four terms. Due to current conservation, the
substitution $P_\sigma \to \Delta^\perp_\sigma/\xi$ may be performed.
It is obvious then that the second term on the r.h.s. of Eq.\ 
(\ref{T-DVCS}) is suppressed by at least one power in $1/{\cal Q}$
w.r.t.\ the leading terms which contain $T_1$ and $T_3$. Once more we
emphasize that $T_3$ is absent at the Born level.

Now we are in a position to discuss the azimuthal angle dependence of the
DVCS and BH amplitudes squared as well as of the interference term. To
ensure that the Fourier series contains no terms which are artificially
generated by kinematical subtleties, we choose a frame rotated w.r.t.\ the
laboratory one\footnote{This reference frame is related to the
centre-of-mass system in Ref.\ \cite{GouDiePirRal97} by a boost of the
hadron in the $z$-direction. }. In our frame the virtual photon has no
transverse components and a negative $z$ component $q_1 = (q_1^0, 0,-|q_1^3|)$. 
The $x$ component of the incoming positron is positive $k = (E, E \sin\theta_e, 
0 ,E \cos\theta_e )$. Other momenta read $P_1 = (M, 0, 0, 0)$ and $P_2 = (E_2, 
|\mbox{\boldmath$P$}_2| \cos\phi \sin\theta_H, |\mbox{\boldmath$P$}_2|
\sin\phi \sin\theta_H, |\mbox{\boldmath$P$}_2| \cos\theta_H)$. $\phi$ is 
the azimuthal angle between the lepton and hadron scattering planes.

The  calculation of the squared amplitudes  is straightforward and
yields (the electric charge is set equal to one):
\begin{itemize}
\item Bethe-Heitler squared term.
\end{itemize}
The amplitude squared for a spin-zero target does not depend on the lepton
polarization. It reads
\begin{equation}
\label{BHsq}
|{\cal T}_{\rm BH}|^2
= \frac{8 F^2}{\Delta^2}
\left\{
\frac{4 M^2 - \Delta^2}{2\Delta^2}
\left(
1 + \frac{{\cal Q}^4 + \Delta^4}
{2(2 k \cdot \Delta -\Delta^2)({\cal Q}^2 + 2 k \cdot \Delta)}
\right)
+ \frac{((q - k)\cdot P)^2 + (k \cdot P)^2}
{(2 k \cdot \Delta -\Delta^2)({\cal Q}^2 + 2 k \cdot \Delta)}
\right\}.
\end{equation}
The denominators of both lepton propagators $(2 k\cdot \Delta
-\Delta^2)$ and $(2 k\cdot\Delta +Q^2)$ depend on the azimuthal angle
$\phi$. Moreover the Taylor expansion of $(Q^2+2 k\cdot\Delta) =
(k-q_2)^2= -\frac{1-y}{y} Q^2 (1+{\cal O}(1/{\cal Q}))$ induces a pole for $y
\rightarrow 1$. We introduce as a new notation
${\cal P}_1$ and ${\cal P}_2$ for the dimensionless lepton propagators:
\begin{equation}
\label{ExaBHpro}
{\cal Q}^2 {\cal P}_1 \equiv (k - q_2)^2 = {\cal Q}^2 + 2k\cdot \Delta,
\qquad
{\cal Q}^2 {\cal P}_2 \equiv (k - \Delta)^2 = - 2 k \cdot \Delta + \Delta^2.
\end{equation}
Their $\phi$ dependence is contained in
\begin{equation}
\label{kDelta}
k\cdot \Delta
= - \frac{{\cal Q}^2}{2y (1 + \epsilon^2)}
\Bigg\{
1 + 2 K \cos{\phi} - \frac{\Delta^2}{{\cal Q}^2}
\left( 1 - \Bx (2-y) + \frac{y \epsilon^2}{2} \right)
+  \frac{y \epsilon^2}{2}
\Bigg\},
\end{equation}
where $\epsilon \equiv 2\Bx M/{\cal Q}$. The kinematical factor
\begin{equation}
K = \left[-\frac{\Delta^2}{{\cal Q}^2} (1 - \Bx)
\left( 1 - y - \frac{y^2\epsilon^2}{4} \right)
\left( 1 - \frac{\Delta^2_{\rm min}}{\Delta^2} \right)
\left\{
\sqrt{1 + \epsilon^2} - \frac{4\Bx (1-\Bx) + \epsilon^2}{4(1-\Bx)}
\frac{\Delta^2_{\rm min} -\Delta^2}{{\cal Q}^2}
\right\}\right]^\frac{1}{2} ,
\end{equation}
is $1/Q$ power suppressed. It vanishes at the kinematical boundary
$\Delta^2 = \Delta_{\rm min}^2$ determined by
\begin{eqnarray}
-\Delta_{\rm min}^2
=  {\cal Q}^2
\frac{2(1 - \Bx) \left(1 - \sqrt{1 + \epsilon^2}\right) + \epsilon^2}
{4\Bx (1 - \Bx) + \epsilon^2}
= \frac{M^2 \Bx^2}{1 - \Bx + \Bx M^2/{\cal Q}^2}
\left\{ 1 + {\cal O} \left( M^2/{\cal Q}^2 \right) \right\} .
\end{eqnarray}
Let us take a closer look at equation\ (\ref{BHsq}). In the second
numerator, $((q-k)\cdot P)^2+(k\cdot P)^2 = (q\cdot P - k\cdot\Delta -2
k\cdot P_1)^2+(k\cdot\Delta+2 k\cdot P_1)^2$, only $k\cdot \Delta$
depends on $\phi$. Therefore equation (\ref{BHsq}) contains terms
proportional to $\cos^0\phi$, $\cos^1\phi$ and $\cos^2\phi$. In other
words (\ref{BHsq}) may be written as
\begin{eqnarray}
\label{Par-BH}
|{\cal T}_{\rm BH}|^2
= -
\frac{F^2(\Delta^2)}{\Bx^2 y^2 (1 + \epsilon^2)^2 \Delta^2\, {\cal P}_1 {\cal P}_2}
\sum_{m = 0}^2 c^{\rm BH}_m \, K^m \cos{(m\phi)}   ,
\end{eqnarray}
where the expansion coefficients are given by
\begin{eqnarray}
c^{\rm BH}_0 \!\!\!&=&\!\!\!
\left\{ (2 - y)^2 + y^2 ( 1 + \epsilon^2 )^2 \right\}
\left\{
4 \Bx^2 \frac{M^2}{\Delta^2} + 4 (1 - \Bx) + (4 \Bx + \epsilon^2)
\frac{\Delta^2}{{\cal Q}^2}
\right\}
\nonumber\\
&&
+ 32\, \Bx^2 K^2 \frac{M^2}{\Delta^2}
+ 2 \epsilon^2
\left\{
4 (1 - y) (3 + 2 \epsilon^2) + y^2 (2 - \epsilon^4)
\right\}
- 4 \Bx^2 (2 - y)^2 (2 + \epsilon^2) \frac{\Delta^2}{{\cal Q}^2},
\nonumber\\
c^{\rm BH}_1 \!\!\!&=&\!\!\!
-8 (2 - y)
\left\{ 2 \Bx + \epsilon^2 - 4 \Bx^2 \frac{M^2}{\Delta^2} \right\},
\nonumber\\
c^{\rm BH}_2 \!\!\!&=&\!\!\!
32 \, \Bx^2 \frac{M^2 }{\Delta^2} \ .
\end{eqnarray}
It should be noted, that all coefficients are bounded in the whole
kinematical region, since
$\frac{M^2}{\Delta^2}$ is always multiplied by $\Bx^2$ and
$\Delta_{\rm min}^2 \propto M^2 \Bx^2$.  The $K$ term in the Fourier
coefficients $c_0$, $c_1$ and $c_2$ suppresses the angular moments by
$\int d\phi \cos{(m\phi)}| {\cal T}_{\rm BH}|^2 \propto \Big\{ -
\frac{\Delta^2}{{\cal Q}^2} \Big( 1 - \frac{\Delta^2_{\rm min}}{\Delta^2} 
\Big) \Big\}^{m/2}$. To prevent induced azimuthal angle dependence due 
to the expansion of lepton propagator in $1/{\cal Q}$, angular
moments with an aditional weight of the two propagators ${\cal P}_1
{\cal P}_2$ can be calculated. Then the averaged BH squared term reads for
large ${\cal Q}^2$
\begin{eqnarray}
\label{Zer-Mom-BH2}
\int_0^{2 \pi} \frac{d\phi}{2\pi}
 {\cal P}_1(\phi) {\cal P}_2(\phi)|{\cal T}_{\rm BH}|^2
= - \frac{F^2(\Delta^2)}{\Bx^2 y^2 \Delta^2}
\left\{
8 (2 - 2 y + y^2) (1 - \Bx)
\left( 1 - \frac{\Delta^2_{\rm min}}{\Delta^2} \right)
+ {\cal O} \left( \frac{1}{{\cal Q}^2} \right) \right\} .
\end{eqnarray}

\begin{itemize}
\item Bethe-Heitler--DVCS interference term.
\end{itemize}
The interference term can be treated in the same way.
\begin{eqnarray}
\label{Def-Mom-Int}
{\cal I} = - \frac{F(\Delta^2)}{\Delta^2\, \Bx^2 y^3 {\cal P}_1 {\cal P}_2}
\left\{ \frac{\Delta^2}{{\cal Q}^2} c^{\cal I}_0 +
\sum_{m=1}^2 K^m \left[c^{\cal I}_m\, \cos{(m\phi)} +
\lambda s^{\cal I}_m\, \sin{(m\phi)} \right] +
\frac{{\cal Q}^2}{M^2} K^3 c^{\cal I}_3\, \cos{(3 \phi)}
\right\}
\end{eqnarray}
As we see the Fourier sums terminate with $\cos{(3\phi)}$ and
$\sin{(2\phi)}$ terms for unpolarized and polarized scattering,
respectively. The information about GPDs is contained in the
dimensionless coefficient functions $c^{\cal I}_m$ and $s^{\cal I}_m$. 
They are linear combinations of the amplitudes $T_i$
introduced in (\ref{T-DVCS}). The exact results are rather
lengthy. Therefore we neglect terms which are power suppressed w.r.t.\ the
leading contribution:
\begin{eqnarray}
\label{AngCoeInt}
c^{\cal I}_0 \!\!\!&=&\!\!\!
- 8 \Bx (2 - y) 
\left\{ (2 - \Bx)(1 - y) - (1 -\Bx)(2 - y)^2
\left(1 - \frac{\Delta^2_{\rm min}}{\Delta^2}\right)
\right\} \, {\rm Re} \, T_1 ,
\nonumber\\
c^{\cal I}_1 \!\!\!&=&\!\!\!
- 8 \Bx (2 - 2y + y^2) \, {\rm Re} \, T_1 ,
\nonumber\\
c^{\cal I}_2 \!\!\!&=&\!\!\!
- 16 (2 - y) \, {\rm Re} \,
\left\{ \Bx T_1 -  T_2 
+ \frac{(2 - \Bx) \Delta^2 -6(1-\Bx) (\Delta^2-\Delta_{\rm min}^2)}
{2(2-\Bx) M^2 } T_3 \right\} ,
\nonumber\\
c^{\cal I}_3 \!\!\!&=&\!\!\!
-\frac{16}{2 - \Bx} {\rm Re} \, T_3 ,
\nonumber\\
s^{\cal I}_1 \!\!\!&=&\!\!\!
8 \Bx y (2 - y) \, {\rm Im} \, T_1 ,
\nonumber\\
s^{\cal I}_2 \!\!\!&=&\!\!\!
16 y \, {\rm Im} \, \left\{ \Bx T_1 - T_2 
- \frac{\Delta^2}{2 M^2} T_3 \right\} .
\end{eqnarray}
As we can see, the interference term falls off like $1/\sqrt{-\Delta^2
{\cal Q}^2}$ for large ${\cal Q}$. Its average over $\phi$ is
suppressed by $1/{\cal Q}^2$. Note that the coefficient $c^{\cal I}_3$
arises at twist-two level due to the gluon transversity in $T_3$.

\begin{itemize}
\item DVCS squared term.
\end{itemize}
Finally, the DVCS squared term reads
\begin{eqnarray}
\label{Def-Mom-DVCS2}
|{\cal T}_{\rm DVCS}|^2 =  \frac{1}{\Bx y^2 {\cal Q}^2}
\left\{ c^{\rm DVCS}_0 + K\left[
c^{\rm DVCS}_1\, \cos{(\phi)}+ \lambda\, s^{\rm DVCS}_1\,\sin{(\phi)}
\right] + \frac{{\cal Q}^2}{M^2} K^2 c^{\rm DVCS}_2\, \cos{(2\phi)}
\right\} .
\end{eqnarray}
The Fourier expansion terminates with $\cos{(2\phi)}$ and has only a
$\sin{(\phi)}$ term. The coefficients are quadratic in the
Compton form factors $T_i$ and to leading order in $1/{\cal Q}$
they are given by
\begin{eqnarray}
\label{AngCoeDVCS}
c^{\rm DVCS}_0 \!\!\!&=&\!\!\!  2(2 - 2y + y^2) \,
\left\{
\Bx T_1 T_1^{\ast}
+ \frac{(1-\Bx)^2}{\Bx (2 - \Bx)^2} 
\frac{(\Delta^2 - \Delta^2_{\rm min})^2}{M^4} T_3 T_3^{\ast} 
\right\},
\\
c^{\rm DVCS}_1 \!\!\!&=&\!\!\! 8 (2 - y)
{\rm Re}\Bigg\{
\left( T_1 - \frac{1- \Bx}{2 - \Bx}
\frac{\Delta^2 - \Delta^2_{\rm min}}{\Bx M^2} T_3 \right)
\left( \Bx T_1 - T_2
- \frac{1 - \Bx}{2 - \Bx}
\frac{\Delta^2 - \Delta^2_{\rm min}}{M^2} T_3 \right)^\ast
\Bigg\},
\nonumber\\
c^{\rm DVCS}_2 \!\!\!&=&\!\!\!
\frac{8}{2 - \Bx}\, {\rm Re} \, T_1 T_3^{\ast} ,
\nonumber\\
s^{\rm DVCS}_1 \!\!\!&=&\!\!\!
-8  y \, {\rm Im}
\Bigg\{
\left( 
T_1 - \frac{1 - \Bx}{2 - \Bx}
\frac{\Delta^2 -\Delta^2_{\rm min}}{\Bx  M^2} T_3 
\right)
\left(
\Bx T_1 - T_2 - \frac{1-\Bx}{2 - \Bx} 
\frac{\Delta^2 -\Delta^2_{\rm min}}{M^2} T_3
\right)^\ast
\Bigg\} .
\nonumber
\end{eqnarray}
The Compton form factor $T_3$ contributes to $c^{\rm DVCS}_0$ and
$c^{\rm DVCS}_2$ at leading twist. In the remaining two coefficients
the interference of $T_1$ with $T_3$ (as well as in $c^{\cal I}_2$ and
$s^{\cal I}_2$) requires a twist-three analysis at next-to-leading
order.  Note that {\it all} angular coefficients are determined by
geometrical (i.e.\ dimension minus spin) twist-two (for $c^{\cal I}_0$,
$c^{\cal I}_1$, $s^{\cal I}_1$, $c^{\cal I}_3$, $c^{\rm DVCS}_0$ and
$c^{\rm DVCS}_2$) or by geometrical twist-three contributions. However,
$c^{\cal I}_3$ and $c^{\rm DVCS}_2$ may be stronger contaminated
by power suppressed contributions than the other coefficients,
since gluon transversity is perturbatively suppressed.

It is important to note, that some of the coefficients are related to
each other. We recall that there are only three Compton form factors 
in equation (\ref{T-DVCS}). Therefore, the four coefficients $c^{\cal I}_i$ 
are reduced to three independent ones. To leading order approximation 
in $1/{\cal Q}$ this relation looks like:
\begin{eqnarray}
\label{Con-Int}
c^{\cal I}_0 =
\frac{2 - y}{2 - 2y + y^2}
\left\{ (2 - \Bx)(1 - y) - (1 -\Bx)(2 - y)^2
\left(1 - \frac{\Delta^2_{\rm min}}{\Delta^2}\right)
\right\} c^{\cal I}_1  + {\cal O} \left(1/{\cal Q}^2 \right).
\end{eqnarray} 
Measuring charge asymmetries yields information about the real part of
all three amplitudes and the imaginary part of two of them (see
discussion below). If we assume for a moment that the $T_3$
contribution is negligibly small, we find the following relations
\begin{eqnarray}
\label{Con-DVCS2}
c^{\rm DVCS}_0 \!\!\!&=&\!\!\! \frac{2 - 2y + y^2}{32 \Bx}  \left\{
\frac{\left(c^{\cal I}_1\right)^2}{(2-2y +y^2)^2} +
\frac{\left(s^{\cal I}_1\right)^2}{y^2 (2-y)^2}
\right\}+ {\cal O} \left(1/{\cal Q}^2 \right),
\nonumber\\
c^{\rm DVCS}_1\!\!\!&=&\!\!\! \frac{1}{16 \Bx} \left\{
\frac{c^{\cal I}_1 c^{\cal I}_2}{2-2y +y^2} +
\frac{s^{\cal I}_1 s^{\cal I}_2}{y^2}
\right\}+ {\cal O} \left(1/{\cal Q}^2 \right),
\\
s^{\rm DVCS}_1\!\!\!&=&\!\!\!\frac{1}{16 \Bx} \left\{
\frac{s^{\cal I}_1 c^{\cal I}_2}{(2-y)^2}-
\frac{c^{\cal I}_1 s^{\cal I}_2}{2-2y +y^2}
\right\}+ {\cal O} \left(1/{\cal Q}^2 \right).
\nonumber
\end{eqnarray}
These constraints can be used to test ${\cal Q}$ scaling of the
Compton form factors. They can be generalized to the case of non-zero
$T_3$. The real part of $T_3$ enters in $c^{\cal I}_3$, while the
imaginary part of $T_3$ can be extracted from $c^{\rm DVCS}_2$
(assuming we know $T_1$). In this case we obtain the same set of
equations (\ref{Con-DVCS2}), but the coefficients $c^{\rm DVCS}_0$,
$c^{\rm DVCS}_1$ and $s^{\rm DVCS}_1$ then depend on all angular moments of
the interference term as well as on $c^{\rm DVCS}_2$.

Finally, let us give a few remarks concerning the actual measurements of the
angular coefficients. Of course, experiments with pions are improbable at
present but the remarks also apply to the analogous case of lepton-nucleon
scattering. For optimal experiments one needs both polarized electron and
positron beams. Since the amplitudes squared are charge-even, while the
interference term is charge-odd, this allows to access all moments of the
interference term. The $\cos{(\phi)}$ and $\sin{(\phi)}$ dependence should
dominate away from the kinematical boundary, because other contributions are
either kinematically or perturbatively suppressed. After subtraction of the
BH squared contribution measurements of the charge-even part of the cross
section for unpolarized lepton beams allow to access all azimuthal angular
moments of the DVCS squared term. Here the constant term is dominant,
because $K$ suppresses the other terms kinematically and because of the
expected smallness of transversity. With a polarized beam the $\sin(\phi)$
term can be directly accessed, because the BH process is independent of the
beam polarization. Following the outlined schedule one should, thus, be able
to test the three equalities (\ref{Con-DVCS2}).

Let us now comment on experiments where a polarized lepton beam is
only available for one charge. Then single spin asymmetries may be
used to yield information about the angular coefficients by weighting
the differential cross section with ${\cal P}_1 {\cal P}_2$.
Unfortunately the DVCS squared term will inevitably contribute to the
azimuthal asymmetries then. Although its contribution is suppressed by
$y\Delta^2/{\cal Q}^2$ [compare Eq.\ (\ref{Def-Mom-Int}) with Eq.\ 
(\ref{Def-Mom-DVCS2})] compared to the interference term it is also
affected by the ratios ${\rm Re} \, T_i/F$. For the unpolarized case
the following observations are relevant. Near the kinematical boundary
$\Delta^2 = \Delta^2_{\rm min}$ all terms of the amplitude squared tend to
be of the same order starting with $1/{\cal Q}^2$. Away from
$\Delta^2=\Delta^2_{\rm min}$ one still has a relative kinematical
suppression of $|{\cal T}_{\rm DVCS}|^2$ by $y\Delta^2/{\cal Q}^2$.
However, $y \Delta^2/{\cal Q}^2$ is now multiplied by terms containing
${\rm Im} \, T_i/F$. Of course, one has to subtract the BH processes
to get the desired information.  Fortunately, for the $\cos(\phi)$ and
$\cos(2\phi)$ moments the BH term enters with the same power $1/{\cal Q}$ 
as the interference term [compare Eq.\ (\ref{Par-BH}) with Eq.\ 
(\ref{Def-Mom-Int})]. The latter again includes the ratios ${\rm Re} \,
T_i/F$.

\section{Twist-three GPDs.}
\label{Sec-Pre}

In this Section we derive the explicit expression for the Fourier
coefficients in terms of GPDs from the hadronic tensor, which is known to
twist-three accuracy in the Born approximation. We will discuss different
aspects of the result. Since factorization has not yet been proven for
$1/{\cal Q}$ suppressed contributions it might happen (in contrast to the
leading twist situation \cite{Rad97,JiOsb98,ColFre98}) that the results are
plagued by singularities. Universality of higher twist distributions could
not be guaranteed then. A keyrole in factorization plays the property that
non-perturbative functions do have specific analytical properties. For
instance, in the case of the pion transition form factor measured in
$\gamma^\ast \gamma^\ast \to \pi^0$, it requires that the meson distribution
should vanish at the end-points and this can indeed be proven
\cite{BroLep80}. In the case of DVCS an analogous requirement is the
continuity of skewed parton distributions at $|x| = \eta$. The tree level
analysis has been recently done \cite{AniPirTer00}-\cite{RadWei00}. For
transverse polarization of the initial photon a divergent amplitude is
encountered. However, this divergence does not show up in the cross section
because of the final state $\gamma$-quantum $\varepsilon^\ast \cdot q_2 = 0$
\cite{RadWei00}. Note that this cancelation of singularities has not yet
been proven for twist contributions higher than twist-three.

For any target the hadronic tensor reads ($\epsilon^{0123} = 1$) in the
considered order \cite{BelMul00}:
\begin{eqnarray}
\label{Tw3Tamplitude}
T_{\mu\nu} (q, P, \Delta)=
- {\cal P}_{\mu\sigma} g_{\sigma\tau} {\cal P}_{\tau\nu}
\frac{q \cdot V_1}{P \cdot q}
+ \left( {\cal P}_{\mu\sigma} P_\sigma  {\cal P}_{\rho\nu}
+ {\cal P}_{\mu\rho}  P_\sigma {\cal P}_{\sigma\nu} \right)
\frac{V_{2 \, \rho}}{P \cdot q}
- {\cal P}_{\mu\sigma} i\epsilon_{\sigma \tau q \rho} {\cal P}_{\tau\nu}
\frac{A_{1\, \rho}}{P \cdot q}.
\end{eqnarray}
$V_{i\rho}$ and $A_{1\rho}$ are given as matrix elements of vector
and axial-vector light-ray operators, respectively.  We should note
that the gauge invariance beyond the twist-three accuracy has been 
restored by hands. Furthermore, in our approximation the vector 
$V_{2 \, \rho}$ depends on the other ones by the relation:
\begin{eqnarray}
\label{V2}
V_{2 \, \rho} = \xi V_{1 \, \rho} - \frac{\xi}{2}
\frac{P_\rho}{P\cdot q} q\cdot V_{1} + \frac{i}{2}
\frac{\epsilon_{\rho\sigma\Delta q}}{P\cdot q} A_{1 \, \sigma}
+ \mbox{twist-four}.
\end{eqnarray}
The Lorentz structure of the $A_{1 \, \sigma}$ term in  (\ref{V2})
when combined with the last term in  (\ref{Tw3Tamplitude}) changes the
gauge invariant ${\cal P}_{\mu\sigma} \epsilon_{\sigma\tau q \rho}
{\cal P}_{\tau\nu}$ projector into $\left( g_{\mu \sigma} - P_\mu \,
q_{2 \sigma}/P \cdot q_2 \right) \epsilon_{\sigma \tau q \rho}
\left( g_{\nu \tau} - P_\nu \, q_{1 \tau}/ P \cdot q_1 \right)$
\cite{BelMul00}. The result (\ref{Tw3Tamplitude}-\ref{V2}) contains
relations that are well known from deep-inelastic scattering.
Obviously, in the forward case, where $V_{i \, \rho} \propto P_{\rho}
F_i$ and $\xi = x_{\rm B}$, we rediscover the Callan-Gross relation.
Besides the generalization of this relation, there are other ones at
twist-three level. They will be discussed below. In our case of a
scalar target, the matrix element of the axial-vector operator has a
non-zero form factor at twist-three level, while the vector operator
already has a non-zero form factor at leading twist level:
\begin{eqnarray}
\label{ParOpeMat}
V_{1 \, \rho} = P_\rho {\cal H} +
(\Delta_\rho -\eta P_\rho)  {\cal H}^{\rm tw3} 
+ \mbox{twist-four},
\qquad
A_{1 \, \rho} = \frac{i\epsilon_{\rho\Delta P q}}{P\cdot q}\;
{\widetilde{\cal H}}^{\rm tw3}.
\end{eqnarray}
The (twist-two and three) form factors ${\cal H}$ and $\widetilde{\cal H}$ 
are given as convolutions in momentum fraction $x$ of perturbatively
calculable hard scattering parts with GPDs ($\otimes \equiv \int dx$):
\begin{eqnarray}
\label{Def-calH}
\left\{
\begin{array}{c}
{\cal H}
\\
\widetilde{\cal H}
\end{array}
\right\}
(\xi, \eta, \Delta^2 | {\cal Q}^2)
=
\sum_{i = u, d, \dots}
\left\{
\begin{array}{c}
C^{(-)}_i
\\
C^{(+)}_i
\end{array}
\right\}
(\xi, x | {\cal Q}^2/\mu^2)
\otimes
\left\{
\begin{array}{c}
H_i \\
\widetilde H_i
\end{array}
\right\}
(x, \eta, \Delta^2 | \mu^2) .
\end{eqnarray}
The summation runs over different parton species and $\mu^2$ is the
factorization scale. To simplify notations we have kept only scaling 
variables as an argument of the functions. For a $Q_i$-charged
quark the leading order coefficient function reads $\xi\,
{C_i^{(\mp)}} \left( \xi, x \right) = Q^2_i (1 - x/\xi - i
\epsilon)^{-1} \mp (x \to -x)$ with $-$ ($+$) standing for parity even
(odd).

The complete twist decomposition of the vector and axial-vector operator is
given in \cite{BelMul00} and our parameterization (\ref{ParOpeMat}) reads
for the twist-two GPDs
\begin{eqnarray}
\label{Def-GPD}
\left( H , \widetilde H \right) (x, \eta) = \int \frac{d\kappa}{2\pi}
{\rm e}^{i \kappa P_+ x}
\langle P_2| {^{V, A}{\cal O}_+} (\kappa, -\kappa) |P_1 \rangle ,
\quad
{^{V, A}{\cal O}_+} (\kappa, -\kappa)
= \bar\psi (- \kappa n)
\gamma_+ \left( 1, \gamma_5 \right) \psi (\kappa n) .
\end{eqnarray}
We omitted gauge links here. Note that these functions can be related
to other fundamental non-perturbative amplitudes, namely hadronic wave
functions, via overlap integrals \cite{LCWrep}. The twist-three
functions contain two pieces, the so-called Wandzura-Wilczek (WW)
piece expressed in terms of the twist-two function and a correlation
function, $G$, of antiquark-gluon-quark operators \cite{BelMul00}:
\begin{eqnarray}
\label{Res-Spi-O}
H^{\rm tw3}(x,\eta) \!\!\!&=&\!\!\!
\int_{-1}^{1}\frac{dy}{|\eta|}
W_+ \! \left( \frac{x}{\eta}, \frac{y}{\eta} \right)
\left(
\stackrel{\rightarrow}{\frac{\partial}{\partial\eta}}
- \frac{y}{\eta} \stackrel{\leftarrow}{\frac{\partial}{\partial y}}
\right) H (y,\eta)  + \frac{1}{\eta} \ H (x,\eta)
- G^{+} (x, \eta) , \nonumber\\
{\widetilde H}^{\rm tw3}(x,\eta) \!\!\!&=&\!\!\!
\int_{-1}^{1} \frac{dy}{|\eta|}
W_-\! \left( \frac{x}{\eta}, \frac{y}{\eta} \right)
\left(
\stackrel{\rightarrow}{\frac{\partial}{\partial\eta}}
- \frac{y}{\eta} \stackrel{\leftarrow}{\frac{\partial}{\partial y}}
\right) H (y, \eta) - G^{-} (x, \eta) ,
\end{eqnarray}
Here $W_\pm (x, y) = \frac{1}{2} \left\{ W (x, y) \pm W(-x, -y)
\right\}$ and the $W$ kernel reads $W (x, y) = {\mit\Theta}_{11}^0 (1
+ x, x - y)$ with \cite{BelMul97} ${\mit\Theta}_{11}^0 (x, y) =
\int_0^1 d \alpha \delta (x \bar \alpha + y \alpha) = [ \theta (x) -
\theta(y) ]/[x - y]$.  Note that the first moment with respect to $x$
vanishes. Moreover, the time reversal invariance together with hermiticity
tells us that the twist-three GPDs are real valued functions with the
symmetry properties $H^{\rm tw3} (- \eta) = - H^{\rm tw3}
(\eta)$, $\widetilde H^{\rm tw3} (- \eta) = \widetilde H^{\rm tw3} 
(\eta)$.  The only new dynamical information is contained in
the antiquark-gluon-quark GPDs \cite{BelMul00}
\begin{equation}
G^{\pm} (x, \eta)
= \int_{-1}^1 \frac{dy}{|\eta|} \int_{-1}^1 du
\left\{
\frac{1 - u}{2}
W^{\prime\prime} \! \left( \frac{x}{\eta},\frac{y}{\eta} \right)
S^+ (y, u, \eta)
\pm \frac{1 + u}{2}
W^{\prime\prime} \! \left( - \frac{x}{\eta}, - \frac{y}{\eta} \right)
S^- (y, u, \eta)
\right\} .
\end{equation}
We used the convention $W^{\prime\prime} \left( \pm \frac{x}{\eta},
\pm \frac{y}{\eta} \right) \equiv \frac{\partial^2}{\partial y^2}
W\! \left( \pm \frac{x}{\eta}, \pm \frac{y}{\eta} \right)$. Due to
this second derivative the antiquark-gluon-quark operators do not
contribute to the second moment of twist-three GPDs. Since the
operators defined in the parity even and odd sectors can be expresses
through each other,  only two independent functions $\langle
P_2| {{\cal S}^{\pm}_{\rho}} ( \kappa, u \kappa, - \kappa ) |P_1
\rangle = P_+^2 \Delta_{\rho}^\perp \int dx \exp (-i \kappa x P_+)
S^{\pm} ( x, u, \eta )$  appear, where the operators are  ${{\cal 
S}^{\pm}_\rho} (\kappa_1, \kappa_2, \kappa_3) = i {\sl g} \bar\psi
(\kappa_3 n) \left[ \gamma_+ G_{+ \rho} (\kappa_2 n) \pm i \gamma_+
\gamma_5 {\widetilde G}_{+ \rho} (\kappa_2 n) \right] \psi (\kappa_1
n)$.  Note that the twist-three form-factors possess discontinuities
\cite{KivPol,RadWei00} at the points $x=|\eta|$, for instance:
\begin{eqnarray}
\label{jumps}
&&H^{\rm tw3}(\eta + 0,\eta) - H^{\rm tw3}(\eta - 0,\eta)
\\
&&\hspace{2cm} =
\frac{1}{2}{\rm PV}\int_{-1}^1 dy\, \frac{1}{y-\eta}
\left\{
\left(
\stackrel{\rightarrow}{\frac{\partial}{\partial\eta}}
+\frac{y}{\eta} \stackrel{\rightarrow}{\frac{\partial}{\partial y}}
\right) H (y, \eta) -
2\stackrel{\rightarrow}{\frac{\partial^2}{\partial y^2}}
\int_{-1}^1 du (1 + u) S^- (y, u, \eta)
\right\},
\nonumber
\end{eqnarray}
where we assumed that the GPDs vanish at the boundary $x=\pm 1$. The
appearence of discontinuities is a general artefact of the procedure
to separate twist-two and -three contributions and is not solely due to
the WW-approximation.

There is a number of relations between the amplitudes $F_i$ in
(\ref{T-gen}) at the Born level. The first one is a generalization of
the twist-two Callan-Gross relation:
\begin{equation}
\label{sumGPD}
F_1 (\xi, \eta) =
\frac{1}{2 \xi} F_2 (\xi, \eta)
= \sum_{i = u, d, \dots} \int_{-1}^1 dx \ C^{(-)}_i (x, \xi) H_i(x, \eta) .
\end{equation}
This equation can be easily deconvoluted by $\sum_{i}\left[ H_i(x,
\eta) - H_i(-x, \eta) \right] = {\rm Im} \, F_1 (x, \eta)/\pi$.
Moreover, by means of antisymmetry we can continue $F_1 (x, \eta)$ to
negative values of $x$. Thus, we obtain a dispersion relation, $\pi
{\rm Re} \, F_1 (\xi, \eta) = {\rm PV} \int_{-1}^1 \ft{dx}{\xi-x} {\rm
Im} \, F_1 (x, \eta)$, which tests the dominance of perturbative
leading order predictions.  In the WW approximation for twist-three
functions two further relations hold, which express the amplitudes 
$F_3$ and $F_4$ in terms of twist-two GPDs:
\begin{eqnarray}
F_3 (\xi, \eta)
\!\!\!&=&\!\!\! \xi \sum_{i = u, d, \dots} Q^2_i \int_{-1}^1 dx
\Bigg\{
\left(
\frac{{\rm sign}(\eta)}{\eta + x} \ln\frac{\xi + \eta - i 0}{\xi - x - i 0}
+ (\eta \to -\eta)
\right)
\left(
\stackrel{\rightarrow}{\frac{\partial}{\partial\eta}}
- \frac{x}{\eta} \stackrel{\leftarrow}{\frac{\partial}{\partial x}}
\right)
\nonumber\\
&&\hspace{3 cm}+  \frac{2}{\eta (\xi - x - i 0)}
- (x \to -x) \Bigg\} H_i (x, \eta) ,
\nonumber\\
F_4 (\xi, \eta)
\!\!\!&=&\!\!\! \xi \sum_{i = u, d, \dots} Q^2_i \int_{-1}^1 dx
\Bigg\{
\left(
\frac{{\rm sign}(\eta)}{\eta + x} \ln\frac{\xi + \eta - i 0}{\xi - x - i 0}
- (\eta \to -\eta)
\right)
\left(
\stackrel{\rightarrow}{\frac{\partial}{\partial\eta}}
- \frac{x}{\eta} \stackrel{\leftarrow}{\frac{\partial}{\partial x}}
\right)
\nonumber\\
&&\hspace{3 cm} - (x \to -x) \Bigg\} H_i (x, \eta) .
\end{eqnarray}
As we can see, in the kinematics of a two-photon process the WW approximation 
exist, although, the $\ln (\xi\pm\eta)$ terms could provide a numerical 
enhancement. In the case of an outgoing on-shell photon, i.e.\ $\xi = 
- \eta + {\cal O} (\Delta^2/{\cal Q}^2)$, one naively expects a logarithmic 
enhancement (stemming from (\ref{jumps})) due to $\ln(\xi + \eta)\to 
\ln(\Delta^2/{\cal Q}^2)$, which would imply difficulties with factorization. 
However, it has been shown that such contributions cancel in physical
amplitudes \cite{RadWei00,KivPol}, and thus legitimize the WW approximation. 
It is interesting to note that the same situation appears also in the 
antiquark-gluon-quark sector.

The DVCS ($\eta = - \xi$) amplitudes in the tensor decomposition
(\ref{T-DVCS}) are given by
\begin{eqnarray}
\label{TsHs}
T_1 = {\cal H}, \qquad
T_2 = 2 \xi {\cal H} + 2 \xi^2
\left\{
{\cal H}^{\rm tw3} - \widetilde{\cal H}^{\rm tw3}
\right\},
\end{eqnarray}
and the form factor $T_3$ is determined at twist-two level by the gluon 
transversity. All Compton form factors also contain twist-four contributions, 
which have not been computed yet. Obviously, the generalized Callan-Gross 
relation can not be tested in the DVCS process, since it is modified by 
twist-three contributions, see (\ref{TsHs}). The modification is given 
as a difference of ${\cal H}^{\rm tw3}$ and $\widetilde{{\cal H}}^{\rm tw3}$ 
and, thus, ensures the cancellation of the $\ln(\xi +\eta)$ term for the WW 
and the antiquark-gluon-quark contributions:
\begin{eqnarray}
\label{Con-tw3}
\left\{ 
{\cal H}^{\rm tw3} - \widetilde{\cal H}^{\rm tw3} 
\right\} (\xi, -\xi)
\!\!\!&=&\!\!\! - \frac{1}{\xi} {\cal H} (\xi, - \xi) -
\sum_{i = u, d, \dots}  \int_{-1}^1 dx
\frac{Q^2_i}{\xi + x}  \ln\frac{2 \xi}{\xi - x - i 0} 
\nonumber\\
&\times&\!\!\! \Bigg\{
\left(
\stackrel{\rightarrow}{\frac{\partial}{\partial \xi}}
- \frac{x}{\xi} \stackrel{\leftarrow}{\frac{\partial }{\partial x}}
\right) 
\left\{ H_i (x, \xi) - H_i (-x,\xi) \right\}
- \stackrel{\leftarrow}{\frac{\partial^2 }{\partial x^2}} S_i(x,\xi)
\Bigg\} ,
\end{eqnarray}
where $S_i(x,\xi) = \int_{-1}^1 du (1 + u) \left[ S_i^+ (-x,-u,-\xi) -
S_i^- (x, u, -\xi) \right]$. Note that this special combination
(\ref{Con-tw3}) of twist-three amplitudes can be represented as ${\cal
H}^{\rm tw3} - \widetilde{\cal H}^{\rm tw3} = \int \frac{dx}{x}
C^{(-)} (x, \xi) \delta H^{\rm tw3} (x, -\xi)$. It is free from
singularities, since the integrand $\delta H^{\rm tw3} (x, -\xi)= x
H^{\rm tw3} - \xi \widetilde H^{\rm tw3}$ is a continuous function in
the momentum fraction $x$. It remains an open problem if this
cancellation of $\ln(-\Delta^2/{\cal Q}^2)$ terms is ensured by
general principles, or if it is only valid in leading order for the
Wilson coefficients and breaks down once radiative corrections are
accounted for. Unfortunately, contrary to the case of general
kinematics, in the WW approximation for DVCS we can not express $T_2$
in terms of $T_1$. To test the WW approximation one has to use models
for GPDs in order to study their influence on $T_1$ and $T_2$.

The tree-level result for the angular coefficients up to twist-three 
accuracy in leading order of the coupling constant\footnote{The twist two 
contribution including gluonic transversity is known in next-to-leading 
order.} reads for
\begin{itemize}
\item the interference term (\ref{AngCoeInt})
\end{itemize}
\begin{eqnarray}
\label{AngCof-Int}
c^{\cal I}_0 \!\!\!&=&\!\!\!
- 8  (2 - y) \Bx \frac{\Delta^2}{{\cal Q}^2}
\left\{
(2 - \Bx) (1 - y) - (1 - \Bx)(2 - y)^2
\left( 1 - \frac{\Delta^2_{\rm min}}{\Delta^2} \right)
\right\} \, {\rm Re} \, {\cal H} ,
\nonumber\\
c^{\cal I}_1 \!\!\!&=&\!\!\!
- 8 (2 - 2y + y^2)  \Bx \, {\rm Re} \, {\cal H},
\qquad
s^{\cal I}_1 
=8  y (2 - y) \Bx\, {\rm Im} \, {\cal H} ,
\\
c^{\cal I}_2\!\!\!&=&\!\!\!
-16  \frac{(2 - y)\Bx}{2 - \Bx} \, {\rm Re} \, {\cal H}^{\rm eff} ,
\qquad\quad
s^{\cal I}_2 =
16  \frac{y \Bx}{2-\Bx} \, {\rm Im} \, {\cal H}^{\rm eff} ,
\nonumber
\end{eqnarray}
\begin{itemize}
\item the DVCS squared term (\ref{AngCoeDVCS})
\end{itemize}
\begin{eqnarray}
\label{AngCof-DVCS}
c^{\rm DVCS}_0 \!\!\!&=&\!\!\!
2  (2 - 2 y + y^2)\Bx {\cal H} {\cal H}^{\ast} ,
\nonumber\\
c^{\rm DVCS}_1 \!\!\!&=&\!\!\!
8 \frac{(2 - y) \Bx}{2 - \Bx}
\, {\rm Re} \,  {\cal H}^{\rm eff} {\cal H}^\ast ,
\qquad
s^{\rm DVCS}_1 =
 8 \frac{y \Bx} {2 - \Bx} \, {\rm Im} \, {\cal H}^{\rm eff} {\cal H}^\ast ,
\end{eqnarray}
where we defined the `effective' twist-three function 
\begin{eqnarray}
\label{Def-efftw3}
{\cal H}^{\rm eff} = - 2 \xi
\left( 
\frac{1}{1 + \xi} {\cal H} 
+ {\cal H}^{\rm tw3} - \widetilde{\cal H}^{\rm tw3}
\right)
\end{eqnarray}
in such a way, that the remaining $\Bx$ dependence of the twist-three
angular coefficients is the same as for twist-two. 
The coefficients $c^{\cal I}_3$ and $c^{\rm DVCS}_2$ contain, besides the
gluonic twist-two transversity, also geometrical twist-four contributions. In
leading order of perturbation theory the latter can be evaluated from handbag
diagrams with additional transversal gluons and, so-called, cat-ear
diagrams. As we observe the twist-three contribution induces a new $\cos (2
\phi)/ \sin (2 \phi)$ angular dependence of charge asymmetries as well as a
new $\cos (\phi)/ \sin (\phi)$ angular dependence in the DVCS squared term
with coefficients proportional to a `universal' combination of GPDs given
in Eq.\ (\ref{Def-efftw3}). In contrast, the angular independent part of the
interference term, i.e.\ $c^{\cal I}_0$, arises from a pure kinematical
twist-three effect.

\section{Double and skewed parton distributions.}
\label{Sec-DD}

Before we give examples for the shape of twist-three GPDs in the next
section let us discuss the relation between skewed and double
distributions (DDs). The double distributions were originally defined by
\cite{MueRobGeyDitHor94}
\begin{eqnarray}
\label{def-DDwro}
\langle P_2| {^V\!\!{\cal O}_+} (\kappa, -\kappa) |P_1 \rangle
= P_+ \int_{-1}^{1} dy \int_{-1 + |y|}^{1 - |y|} dz\,
{\rm e}^{- i \kappa P_+ y -i \kappa \Delta_+ z} F (y, z).
\end{eqnarray}
By means of the so-called $\alpha$-representation it has been shown
that $F(y, z)$ is a generalized function defined in the region
$|y| + |z| \le 1$. This original definition, also introduced in
\cite{Rad97}, leads to a violation of the polynomiality condition for
GPDs \cite{Ji97}.  The reasoning goes as follows. The $j$th-moment of
an GPD is given by the expectation value of local twist-two operators
with spin $j + 1$, i.e.\ ${^V\!{\cal R}}^2_{\rho; \mu_1 \dots \mu_j} =
{\mathop{\mbox{\large\bf S}}}_{\rho \mu_1 \dots \mu_j} \bar\psi
{\mit\gamma}_\rho \,i\!\stackrel{\leftrightarrow}{\cal D}_{\mu_1}
\dots i\!\stackrel{\leftrightarrow}{\cal D}_{\mu_j}\psi $, which are
completely symmetrized and traceless, and where
$\stackrel{\leftrightarrow}{\cal D}_{\mu} =
\stackrel{\rightarrow}{\cal D}_{\mu} - \stackrel{\leftarrow}{\cal
  D}_{\mu}$ and ${\cal D}_\mu = \partial_\mu- i {\sl g} B_\mu$.  Thus,
they are polynomials of order $j+1$ in $\eta$:
\begin{eqnarray*}
\int_{-1}^{1} dx x^j H(x,\eta) =n_\rho n_{\mu_1}\dots n_{\mu_j}
\langle P_2| {^V\!{\cal R}^2_{\rho; \mu_1 \dots \mu_j}} |P_1 \rangle
= P_+^{j + 1} \sum_{k = 0}^{j + 1} \eta^k H_{j + 1,j + 1 - k}.
\end{eqnarray*}
On the other hand
\begin{equation}
\label{RedFor-wro}
H(x, \eta) = \int_{-1}^{1} dy \int_{-1 + |y|}^{1 - |y|} dz\,
\delta (y + \eta z - x) F(y, z)
= \int_{- 1}^{1} \frac{dy}{\eta} \, {\mit\Xi} (y| x, \eta)
F \left( y, \frac{x - y}{\eta} \right),
\end{equation}
with
\begin{eqnarray}
{\mit\Xi} (y| x, \eta)
= \theta (x > \eta)
\theta \left(
\ft{x + \eta}{1 + \eta} > y > \ft{x - \eta}{1 - \eta}
\right)
\!\!\!&+&\!\!\! \theta (- \eta > x)
\theta \left(
\ft{x + \eta}{1 - \eta} > y > \ft{x - \eta}{1 + \eta}
\right) \nonumber\\
&+&\!\!\! \theta (\eta > x > - \eta)
\theta \left(
\ft{x + \eta}{1 + \eta} > y > \ft{x - \eta}{1 + \eta}
\right) . \nonumber
\end{eqnarray}
The $j$th-moment of $H(x, \eta)$ generates only a polynomial of order
$j$ in $\eta$: $\int dx\, x^j H(x,\eta) = \int dy \int dz\, (y + \eta
z)^j F(y,z)$.  As suggested in Ref.\ \cite{PolWei99} one can cure this
problem by adding in the definition (\ref{def-DDwro}) of DDs an extra
independent term concentrated in $y = 0$ and proportional to
$\Delta_+$:
\begin{eqnarray}
\label{RedFor-PW}
H(x, \eta) =
\int dy \int dz\,
\delta(y + \eta z - x) \left\{ F(y, z)  + \eta \delta(y) D(z) \right\} .
\end{eqnarray}
Since $D(z)$ has the support $|z|\le 1$, it induces a term
in the GPD that is entirely concentrated in the `exclusive' region. Indeed
this choice is unique and the $D$-term can be extracted from a given
distribution by $D(x) = \lim_{\eta \to \infty} H (\eta x, \eta)$. The
first term in the bracket of Eq.\ (\ref{RedFor-PW}) is understood as a
representation of a subtracted GPD, namely, $H(x,\eta) - {\rm sign}(\eta) 
D(x/\eta)$. Note that $D(z)$ is an antisymmetric function in $\eta$ since 
$H (x, - \eta) = H (x, \eta)$.

In the following we give an alternative solution of the problem stated
above, in which the spectral function does not depend on the
skewedness parameter $\eta$. The new representation will enable us to
derive an inverse transformation in a simple way, since it does not
explicitly depend on $\eta$. It is instructive to start with local
operators. The parameterization for the matrix elements of the
completely symmetrized and traceless local vector operators sandwiched
between spin-zero states reads:
\begin{eqnarray*}
\langle P_2| {^V\!{\cal R}^2_{\rho; \mu_1 \dots \mu_j}} |P_1\rangle
\!=\!\! \Sym_{\rho \mu_1 \dots \mu_j}
\!\left\{
P_\rho \dots P_{\mu_j} H_{j+1, j+1}
+ \Delta_\rho P_{\mu_{1}} \dots P_{\mu_{j}} H_{j+1, j}
+ \dots
+ \Delta_\rho\dots \Delta_{\mu_j} H_{j+1, 0}
\right\} .
\end{eqnarray*}
Next we introduce a generating function for the coefficients $H_{jk}$.
It is easy to check that after contraction with the light-cone vector
$n_\mu$ and summation over the local operators, i.e.\
${\cal O}_+(\kappa, - \kappa) = n_\rho {\cal R}^2_\rho (\kappa, - \kappa)$,
${\cal R}^2_\rho (\kappa , - \kappa) = \sum_{j = 0}^{\infty}
\frac{(-i\kappa)^j}{j!} n_{\mu_1} \dots n_{\mu_j} {\cal R}^2_{\rho;
\mu_1 \dots \mu_j}$, the following definition of $H (x,\eta)$
\begin{eqnarray*}
H_{j, j - k} = \frac{1}{k!}
\frac{d^k}{d \eta^k} \int_{-1}^{1} dx\; x^{j - 1}
H(x,\eta)_{|\eta = 0},
\qquad\mbox{where}\qquad
0 \le k \le j, \quad 1 \le j,
\end{eqnarray*}
is equivalent to that one in Eq.\ (\ref{Def-GPD}). Let us remark that
the term proportional to $\Delta_\rho$ in the matrix elements of the
twist-two light-ray operator ${\cal R}^2_\rho (\kappa, - \kappa)$ arises
from combinatorics and is fixed by a WW type relation
\cite{BelMul00}
\begin{eqnarray}
\label{ExpinGPD}
\langle P_2 | {^V\!{\cal R}^2_\rho} (\kappa, - \kappa)| P_1\rangle
= \int_{-1}^1 dx {\rm e}^{-i \kappa P_+ x}
\left(
P_\rho H (x, \eta)
+ \Delta_\rho^\perp \int_{-1}^{1} dy \,
W_2 (x, y) \frac{d}{d\eta} H (y, \eta)
\right),
\end{eqnarray}
where the kernel reads $W_2(x, y) = {\mit\Theta}^0_{11} (x, x - y)$.

Analogous to the definition of the GPD as a generating function of moments
$H_j (\eta)$, we now introduce a double distribution $f (y, z)$ with
the moments $H_{j, j - k} \equiv \left({j \atop k}\right) \int dy \int
dz\;  y^{j-k} z^{k} f (y, z)$ where $0 \le k \le j$, $1 \le j$. Summing
the series of moments up again we find
\begin{eqnarray}
\label{RedFor-tru}
H(x,\eta) = \int dy \int dz\, x \delta(y + \eta z - x) f (y, z).
\end{eqnarray}
so that the transformation (\ref{RedFor-wro}) is modified in a minimal
way by an additional factor of $x$. The matrix elements of twist-two
light-ray operators are expressed in terms of the new double distribution
by
\begin{eqnarray}
\label{Def-cor-DD}
\langle P_2 | {^V\!{\cal R}^2_\rho} (\kappa, - \kappa)| P_1\rangle
= \int dy \int dz f(y,z)  \left(y P_\rho + z \Delta_\rho \right)
{\rm e}^{- i \kappa P_+ y -i \kappa \Delta_+ z}.
\end{eqnarray}

Now we are able to derive an inverse transformation for equation
(\ref{RedFor-tru}). In the first step we project the Lorentz index
$\rho$ in Eq.\ (\ref{Def-cor-DD}) onto the transverse plane and
subsequently perform Fourier transformations w.r.t. $\Delta_+$ and
$\kappa$. Employing the representation (\ref{ExpinGPD}) for the matrix
elements in terms of GPDs, we immediately obtain the desired transformation:
\begin{eqnarray*}
z f(y,z) = -\frac{1}{2\pi^2} \int dx \int d\eta\;
{\rm PV} (y + \eta z - x)^{- 2} 
\int dx^\prime W_2(x,x^\prime) \frac{d}{d\eta} H(x^\prime,\eta).
\end{eqnarray*}
Here we used the following Fourier representation for $|\kappa P_+|$:
$|a| = - \ft{1}{\pi} \int d\lambda\; \exp \left( i \lambda a \right)
{\rm PV} \ft{1}{\lambda^2}$, where we understand the PV prescription
as  ${\rm PV} \ft{1}{\lambda^2} \equiv \frac{1}{2}\left\{\ft{1}{(\lambda+i0)^2}
+\ft{1}{(\lambda-i 0)^2} \right\}$. The final steps are to perform the
$x$ integration and a partial integration w.r.t.\ $\eta$. In the last
step we dropped a surface term. We find
\begin{eqnarray}
\label{Res-Inv}
f(y, z) = \frac{1}{2\pi^2} \int d\eta \int \frac{dx}{x}
{\rm PV} \left\{ (y + z \eta)^{- 2} - (y + z\eta - x)^{- 2} \right\}
H (x, \eta) .
\end{eqnarray}

Finally, we express the $F-$ and $D$-functions of the representation
(\ref{RedFor-PW}) in terms of the $f$-function. The $D$-term is easily
extracted by taking the limit $\eta,x\to \infty$ with $x/\eta$ fixed:
\begin{eqnarray}
D(z) = z \int dy\, f(y,z).
\end{eqnarray}
Consequently, the term $\int dy \int dz\, x \left\{ \delta(y + \eta z - x) 
- \delta(\eta z - x) \right\} f (y, z)$ corresponding to the first
one on the r.h.s.\ of Eq.\ (\ref{RedFor-PW}) can be cast in the form
(\ref{RedFor-PW}) by means of
\begin{eqnarray*}
x \left\{
\delta(y + \eta z - x) - \delta(\eta z - x) \right\}
= y \delta(y + \eta z - x) + z \frac{d}{dz} \int dy^\prime
\delta(y^\prime + \eta z - x) W_2 (y^\prime, y) y .
\end{eqnarray*}
It leads to
\begin{eqnarray}
\label{ConToPW}
F(y,z) = y f(y,z) -
\frac{d}{dz} z \int dy^\prime W_2(y,y^\prime) y^\prime f(y^\prime,z).
\end{eqnarray}
Thus, both $F$ and $D$ functions in (\ref{RedFor-PW}) turn out to be
different projections of the same function $f$.
Note that the projection (\ref{ConToPW}) together with the inversion
formula (\ref{Res-Inv}) give us the function $F(y,z)$ in terms of
the GPD:
\begin{eqnarray}
F(y, z) = -\frac{1}{2\pi^2} \int d\eta \int dx
{\rm PV} \left\{  (y + z\eta - x)^{- 2} \right\}
\left[H(x,\eta) - {\rm sign}(\eta) D(x/\eta)\right].
\end{eqnarray}
Here the $D(x/\eta)$ term appears from a surface term that
arised from the convolution with the $W_2$ kernel [see r.h.s.\ of Eq. (42)].

\section{Features of WW approximation.}
\label{Sec-NumEst}

In this Section, which aims to serve illustrative purposes only, we will
give qualitative estimates of the twist-three contributions. An analysis
and numerical estimates for targets with spin will be given elsewhere.
In the following we assume that the GPD factorizes in a partonic form
factor $F(\Delta^2)$ and a function depending only on the remaining
three variables $H (x, \eta, \Delta^2 | {\cal Q}^2) = F (\Delta^2) H
(x, \eta| {\cal Q}^2)$. The non-trivial polynomiality requirement for
the moments of $H (x, \eta)$ is obeyed by using the transformation
formula discussed in the previous Section. To ensure the reduction
of the GPD to the parton densities in the forward case, we assume that
the double distribution can be modeled, following the proposal in Ref.\ 
\cite{Rad99}, as a product of the quark densities with a
profile function that has certain properties. In the case of the
transformation (\ref{RedFor-PW}) the ansatz for the quark DD is ($\bar
y \equiv 1 - y$)
\begin{eqnarray}
F_i (y, z) = f_i ( y ) \pi (|y|, z),
\quad
\pi (y,z) = \frac{3}{4} \frac{ \bar y^2 - z^2 }{ \bar y^3},
\quad
f_i (y) = q_i (y) \theta (y) - \bar q_i (- y) \theta (- y) .
\end{eqnarray}
In the following we limit ourselves to the valence quark approximation,
thus, we set $\bar q = 0$ and $q = q_{\rm val}$ which is, for the sake of
simplicity, chosen to be determined by Regge and quark counting rules
$q_{\rm val} (y) = q_u + q_{\bar d} = \frac{3}{2} y^{-1/2} (1 - y)$.

\begin{figure}[t]
\vspace{-0.5cm}
\begin{center}
\mbox{
\begin{picture}(0,140)(300,0)
\put(60,0){\insertfig{8}{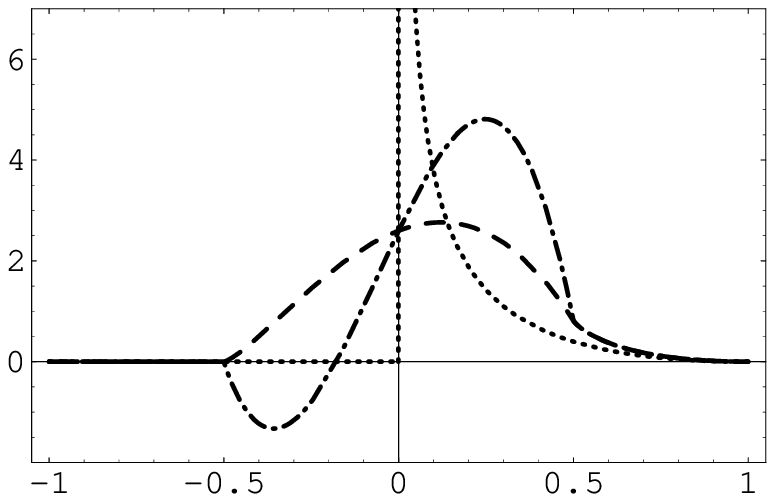}}
\put(100,110){(a)}
\put(280,-10){$x$}
\put(300,0){\insertfig{8.2}{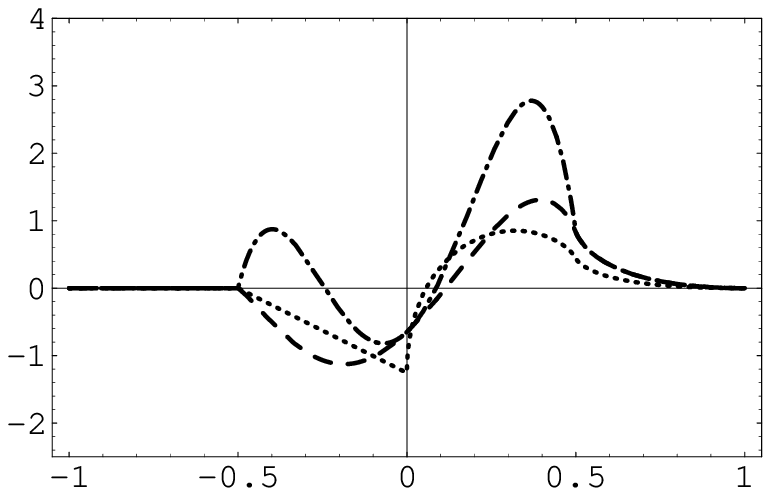}}
\put(335,110){(b)}
\put(530,-10){$x$}
\end{picture}
}
\end{center}
\caption{\label{GPDs&WW}  Twist-two GPDs $H(x, \eta)$ $\eta = \ft12$
are shown in (a) for the
transformation (\ref{RedFor-PW}) without (dashed) and with (dash-dotted)
$D$ term and the FPD model (dotted). In (b) we plot the corresponding
WW approximation for $\delta H ^{\rm tw3} (x, \eta)$ for $\eta = \ft12$. 
Changing $\eta$ to $- \xi$ leaves (a) intact and reflects (b) w.r.t.\
the horizontal axis.}
\end{figure}

In Fig.\ \ref{GPDs&WW} (a) the shape of the GPD that results from the 
transformation (\ref{RedFor-PW}) with $D = 0$ is shown (dashed).
Taking a simple ansatz\footnote{The sign and normalization are not relevant
for our present discussion.} for the $D$-term, i.e.\ $D(z) = \theta(1-|z|) 
2 z (1 - z^2)$, we see that such a term induces a complex shape (dash-dotted 
line). We also plotted the so-called forward parton distribution (FPD) 
model, where the GPD is equated to the quark density, i.e. $\pi(y, z) = 
\delta(z)$. As we already know, the WW approximation for the twist-three 
GPDs induces jumps at point $|x|=\xi$. However, they cancel in the 
combination $\delta H^{\rm tw3}$, see Fig.\ \ref{GPDs&WW} (b).

\begin{figure}[t]
\vspace{-1cm}
\begin{center}
\mbox{
\begin{picture}(0,300)(300,0)
\put(60,150){\insertfig{8}{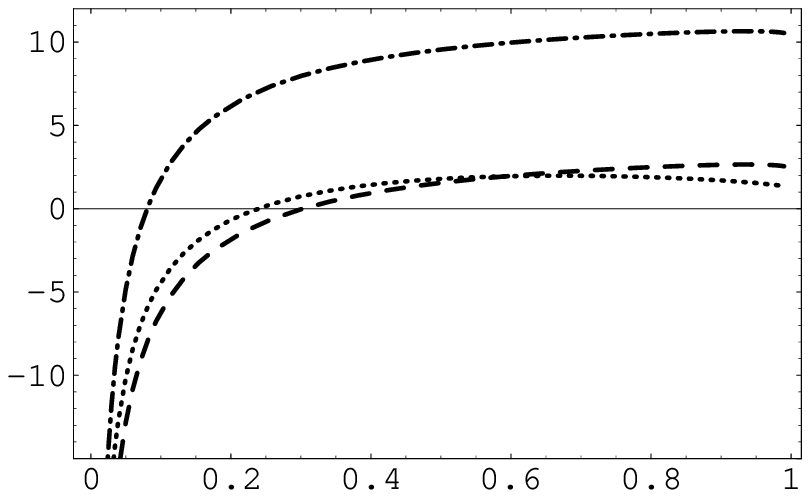}}
\put(230,200){(a)}
\put(470,200){(b)}
\put(300,150){\insertfig{8.25}{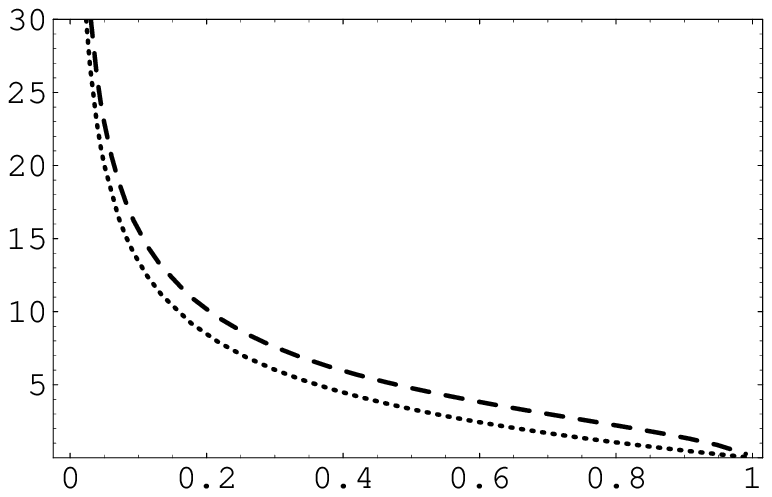}}
\put(60,0){\insertfig{8}{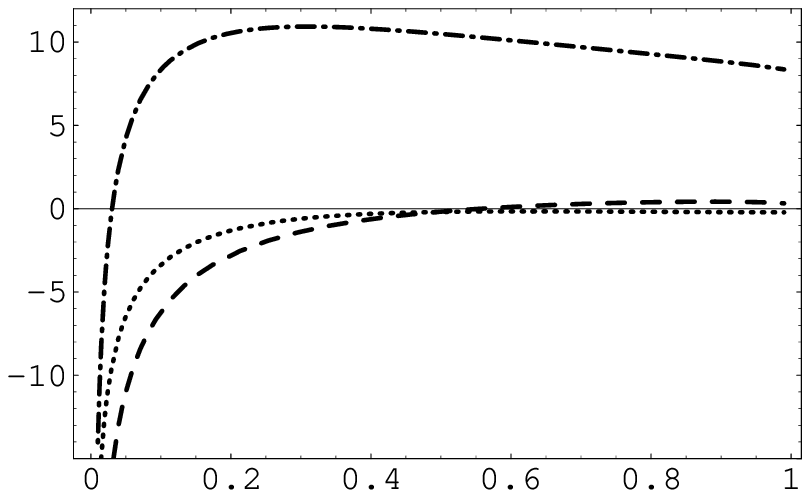}}
\put(300,0){\insertfig{8.25}{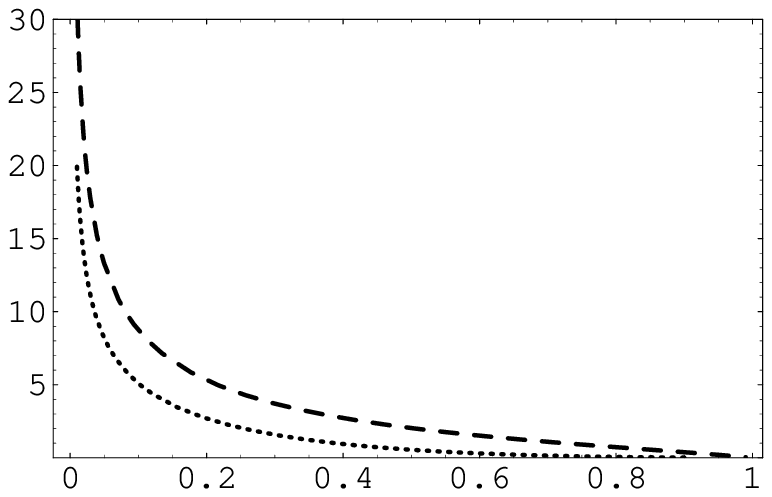}}
\put(230,50){(c)}
\put(470,110){(d)}
\put(280,-10){$\xi$}
\put(530,-10){$\xi$}
\end{picture}
}
\end{center}
\caption{\label{Fig-LOpre} The twist-two predictions (a,b) and the WW 
approximation (c,d) of ${\cal H}^{\rm eff-WW}$, for the convolution of the 
hard scattering amplitude with the GPDs specified in Fig.\ (\ref{GPDs&WW}) 
are shown for the real (a,c) and imaginary part (b,d), respectively.
}
\end{figure}

In Fig.\ \ref{Fig-LOpre} (a,b) we show the DVCS amplitude ${\cal H}$
as defined in Eq.\ (\ref{Def-calH}) for the same twist-two models as
introduced above. Here we observe that the DD-distribution model
(\ref{RedFor-PW}) without $D$-term and the FPD model give almost the
same DVCS amplitude. We also see that a $D$-term can change the real
part drastically, while it drops out in the imaginary part of the
amplitude. It is worth mentioning that quite different shapes of the 
GPD do not change the simple convex (concave) form of the real 
(imaginary) part of the DVCS amplitude.

Now let us address the WW approximation of the `universal' effective 
combination (\ref{Def-efftw3}) that determines the normalization 
of the twist-three angular coefficients (\ref{AngCof-Int}), 
(\ref{AngCof-DVCS}). We decompose it in WW and antiquark-gluon-quark 
pieces ${\cal H}^{\rm eff} (\xi) = {\cal H}^{\rm eff-WW} (\xi) + 
{\cal H}^{\bar q G q} (\xi)$. In the definition (\ref{Con-tw3}) we 
rewrite the derivatives\footnote{Obviously, for any smooth function 
$\tau(z)$, the following identity $\frac{\partial}{\partial x}
\frac{x \tau(x/\xi)}{\xi^2} = - \frac{\partial}{\partial \xi} 
\frac{\tau(x/\xi)}{\xi}$ holds true.} and find 
\begin{eqnarray}
\label{Def-EffWW}
{\cal H}^{\rm eff-WW} = \frac{2}{1 + \xi}{\cal H} 
+ 2 \xi \frac{\partial}{\partial \xi} \sum_{i=u,d,\dots}
\int_{-1}^1 dx \frac{Q^2_i}{\xi + x} \ln\frac{2 \xi}{\xi - x - i 0}
\left\{ H_i(x, \xi) - H_i(- x, \xi) \right\} .
\end{eqnarray}
Since the integrand only has a logarithmic singularity at $x = \xi$,
the integration will result into a smooth $\xi$ dependence, which,
however, will be partly removed by the differentiation w.r.t.\ $\xi$.
As we can see in Fig.\ \ref{Fig-LOpre} (c,d) for the real part the
normalization and shape of ${\cal H}^{\rm eff-WW}$ differs only
slightly from that of ${\cal H}$, while the differences in the
normalization for the imaginary part are caused by the behaviour of
the twist-two GPD at the point $x = \xi$. In any case we observe {\em no} 
numerical enhancement of the twist-three contributions in the WW
approximation. The curves suggest a direct connection of the WW
approximation with the twist-two GPD in the small $\xi$ region. For
the FPD model we can even analytically perform the integration and the
amplitude is entirely determined by the small $x$ behaviour of the
parton density. In the case $q(x) = A x^{-a}$ with $a > 0$ for $x\to
0$, we immediately obtain the imaginary and real parts:
\begin{eqnarray}
&&{\rm Im} {\cal H} = A \pi \xi^{-a}, 
\hspace{2.6cm}
{\rm Im} {\cal H}^{\rm eff-WW} = 2 A \pi \xi^{-a}
\left[ 1 - \frac{a}{2} \psi \left( \frac{1 + a}{2} \right) 
+ \frac{a}{2} \psi \left( \frac{a}{2} \right) \right] ,
\nonumber\\
&&{\rm Re} {\cal H} 
= - A \pi \xi^{-a} \cot \left( \frac{a \pi}{2} \right) , 
\quad
{\rm Re} {\cal H}^{\rm eff-WW} = - 2 A \xi^{-a}
\left[ \pi \cot \left( \frac{a \pi}{2} \right) - a C(a) \right],
\end{eqnarray}
where $C(a) = \int_0^\infty dx\, x^{-a} \left[ \ln\left(\frac{|1-x|}{2}
\right)/(1+x) - \ln\left(\frac{1 + x}{2}\right)/(1 - x)\right]$. It is 
important to note that a different ansatz can give quite different 
amplitudes in the small $\xi$ region. This fact may be helpful in order 
to distinguish GPD models by comparison with small $\Bx$ data from 
HERA experiments.

Finally, we take a look beyond the WW approximation. Both the
measurement of the polarized structure functions in deep inelastic
scattering as well as lattice data suggest that in the forward limit
the WW approximation is quantitatively valid. For the DVCS process we
have to compare the size of the effective twist-three contribution
(\ref{Def-EffWW}) in the WW approximation with that one coming from the
antiquark-gluon-quark correlation functions:
\begin{eqnarray*}
{\cal H}^{\bar q G q} (\xi) 
= -2 \xi \sum_{i = u, d, \dots}  \int_{-1}^1 dx
 \frac{Q^2_i}{\xi + x}  \ln\frac{2 \xi}{\xi - x - i 0}
\stackrel{\leftarrow}{\frac{\partial^2 }{\partial x^2}} S_i(x,\xi).
\end{eqnarray*}
Since the coefficient function contains now a double pole at $x=\xi$,
one has in general a numerical enhancement of $S_i(x,\xi)$. Thus, the
antiquark-gluon-quark contribution would have to be much smaller than
$H_i(x,\xi)$ to favour the WW approximation. This can presumably be
tested by extracting different azimuthal angular moments from
experimental data as discussed in the previous Sections.

\section{Conclusions.}
\label{Sec-Con}

In the present contribution we have studied DVCS on a scalar target to
twist-three accuracy. Let us summarize the lessons we have learnt
from this exercise:

\begin{itemize}
  
\item The interference term is dominated by $\cos{\phi}/\sin{\phi}$
  dependence away from the kinematical boundary. ${\cal Q}$ scaling of
  the Compton form factors can also be tested by a set of relations
  among the azimuthal angular coefficients.

\item Twist-three effects do not show up in the coefficients of the
  azimuthal angle dependence already present in the leading twist
  approximation but rather they induce new Fourier components. Thus, we
  expect at most ${\cal O} (\Delta^2/{\cal Q}^2)$ and ${\cal O}
  (M^2/{\cal Q}^2)$ corrections to the leading twist angular
  dependence.
  
\item Twist-three functions contribute in a singularity free
  combination to the physical cross sections and there is no violation
  of factorization.
  
\item GPDs are related to a single spectral density, or double
  distribution.
  
\item Rather different shapes of GPD models result in quite similar
  shapes for the predicted DVCS amplitudes manifesting low sensitivity
  to the former. However, the overall normalization differs
  significantly.

\end{itemize}

A similar analysis for the DVCS cross section on the nucleon will be
given elsewhere.

\vspace{0.5cm}

We are grateful to V.M. Braun, M. Diehl, A. Freund, N.A. Kivel, M.V. Polyakov, 
A.V. Radyushkin and O.V. Teryaev for discussions on different aspects of this
paper.

\vspace{0.5cm}

Note added: Recently it has been pointed out in \cite{Ter01} that the inversion 
of Eq.\ (\ref{RedFor-tru}) for the function $H (x, \eta)/x$ is known as the 
inverse Radon transformation \cite{GelShiVil66}. Our result (\ref{Res-Inv}) 
differs from the standard formula by the first term in the integral on the 
r.h.s.\ of Eq.\ (\ref{Res-Inv}) [and a surface term]. Obviously, our integral 
exist also in the case if $H(x, \eta)/x$ has a non-integrable singularity at 
$x=0$. In the case when $H(x, \eta)/x$ is integrable, the first term in the 
integrand  can be safely neglected, being proportional to $\delta (y) \delta (z)$, 
and we obtain the inverse Radon transformation for the function $H (x, \eta)/x$. 
Note that the PV prescription can be rewritten as ${\rm PV} \int_{-\infty}^\infty 
dz\; \tau(z)/z^2 = \int_{0}^\infty dz \{\tau(z)+\tau(-z) -2\tau(0)\}/z^2$ 
\cite{GelShi64}.

\end{document}